\documentclass[conference]{IEEEconf}

\input epsf
\usepackage{graphicx}
\usepackage{multirow}
\usepackage{titlesec}
\titleformat{\subparagraph}[runin]{\normalfont\normalsize\bfseries}{\thesubparagraph}{1em}{}
\usepackage{balance} 
\usepackage{stfloats}
\usepackage[table]{xcolor}
\usepackage{xcolor}
\usepackage{comment}
\usepackage{float}
\usepackage{svg}
\usepackage{enumitem}
\usepackage{cite}
\usepackage{lipsum}
\usepackage{graphicx}
\usepackage[colorlinks=true,linkcolor=blue,citecolor=blue,urlcolor=blue]{hyperref}

\usepackage[most]{tcolorbox} 
\renewcommand\thesection{\arabic{section}} 

\renewcommand\thesubsectiondis{\thesection.\arabic{subsection}}

\renewcommand\thesubsubsectiondis{\thesubsectiondis.\arabic{subsubsection}}

\renewcommand{\abstractname}{\textbf{Abstract}}

\newtcolorbox{sectionbox}{
  boxsep=0pt,
  left=1pt,
  right=1pt,
  top=2pt,
  bottom=2pt,
  before skip=2pt,
  after skip=2pt,
  halign=center,
  enhanced,
  arc=0.8mm
}

\titleformat{\subparagraph}[runin]{\bfseries}{\thesubparagraph}{1em}{} 
\begin{document}

\title{\textbf{\Large Beyond Models: Reflections on Engineering AI-enabled Systems in
a Project-Based Course\\}}
	
   

\author{Amir Mashmool$^{1,*}$, Kishan Ravindra Sawant$^{1}$\\, Mojtaba Shahin$^{2}$, Nico Hochgeschwender$^{1}$, and Rainer Koschke$^{1}$ \\ \normalsize $^{1}$Department of Mathematics and Computer Science, University of Bremen, Bremen, Germany\\ \normalsize $^{2}$School of Computing Technologies, RMIT University, Melbourne, Victoria, Australia\\ \normalsize \{mashmool.amir, kishan.sawant, nico.hochgeschwender, koschke\}@uni-bremen.de, \\ mojtaba.shahin@rmit.edu.au\\ \normalsize *corresponding author }



\maketitle
\begin{abstract}

\textbf{Teaching Software Engineering for AI-enabled systems entails addressing the integration of AI components within full-scale software architectures under realistic constraints. While machine learning courses emphasize model development, students often lack experience in architectural design, deployment, and monitoring of AI-enabled systems. Empirical evaluations of such system-oriented AI courses remain limited. This paper reflects on the design and implementation of a project-based master's-level course titled \textit{AI Algorithms: Theory and Engineering}, at the University of Bremen, in which students developed a movie recommendation system while making architectural design decisions to address challenges related to scalability, deployment, and evolving requirements. We conducted a mixed-methods study combining analyses of student submissions and questionnaire responses to investigate integration challenges, learning outcomes, and opportunities for improvement. 
Our results indicate persistent difficulties in early architectural decisions, heterogeneous ML integration, evolving requirements, and data management, largely due to uneven ML and software engineering expertise. From the educator’s perspective, the course fostered system-level reasoning and strengthened awareness of data-centric ML practices in AI-enabled systems.}

\end{abstract}
\IEEEoverridecommandlockouts
\vspace{1.5ex}

%
\IEEEpeerreviewmaketitle

\section{Introduction}
\label{Introduction}

Artificial intelligence (AI) capabilities are increasingly embedded in contemporary software systems, enabling adaptivity and data-driven decision-making across domains. However, engineering AI-enabled systems poses challenges beyond the design of learning algorithms and model architectures. As Sculley \emph{et al.}~\cite{sculley2015} argue, in their seminal work on hidden technical debt in machine learning (ML) systems, real-world AI-enabled systems arise not from isolated models but from an entanglement of code, data, configuration, and infrastructure. These insights urge us to not only re-express software engineering (SE) principles for AI, such as emphasizing modularity and abstraction, validation strategies, and lifecycle management, but also to educate future engineers of AI-enabled systems to understand and apply them in a systematic manner to engineer high-quality systems.

Graduate students often lack hands-on experience in software engineering. To bridge the gap between academia and industry, institutions should integrate project-based courses into their curricula. This paper reflects on our project-based course designed to expose students to real-world challenges in engineering AI-enabled systems and addresses the following research questions:

\textbf{RQ1: What are the most common challenges and mistakes students face when integrating ML methods to develop AI-enabled systems?} 

With AI-enabled systems being integrated across domains, students are expected to learn and apply these technologies. However, applying them is challenging due to the expertise required in data handling, model selection and adaptation, evaluation, and deployment.

As many students lack prior experience in developing such systems, exposure to these challenges helps develop relevant skills.

Identifying these common challenges is guide educators to design more effective teaching strategies.

\textbf{RQ1.1: From the educators’ perspective, what are the most challenging and most common mistakes students make in their assignments and final projects?} Based on assignments and project submission, teachers' insights can provide a deeper look into technical quality, design choices, and how well learning goals were achieved.

\textbf{RQ2: How effectively does the course design assist students in moving from theoretical concepts to practical implementation in building a real-world AI-enabled system?}

Integrating practical projects with theoretical courses can bridges the gap between conceptual understanding and the real-world challenges students face. The effective design of projects and assignments is therefore critical in supporting this transition, enabling students to apply theoretical knowledge to concrete, real-world AI-enabled systems.

\textbf{RQ3: What are the takeaways of students from the course in the perspective of software engineering, ML, and their integration?} While assignments and projects provide objective evidence of student performance, students' perceptions and reflections provide valuable insight into what they learned, the learning process, and their understanding of the integration of software engineering and ML.

\textbf{RQ4: How can the course be further improved by looking into project submissions and questionnaire responses?} 

Identifying opportunities for adapting course design strategies allows educators to build on first-hand experience. The common challenges and mistakes observed in students’ work, together with the extent to which learning outcomes were achieved, motivate the exploration of possible improvements and restructuring of the course.

To address the above research questions, we conducted a mixed-methods study that combines an analysis of students’ project submissions and their responses to a survey. The students’ project submissions provide concrete evidence of their understanding and application of course concepts as well as insights into the challenges they faced. In addition, survey responses capture their perceptions of these challenges and their suggestions for improvement. Together, these data sources help identify recurring issues, gaps in learning, and opportunities for refining the course design, assignments, and project structure.

The key contributions of this paper are:
\begin{itemize}[left=0pt]

\item The design and implementation of a project-based master’s-level course operationalized via five scaffolded assignments leading to an end-to-end AI-enabled system, that explicitly addresses the gap between ML theory and system-level software engineering practice.
\item An empirical analysis of student assignments, final projects, and survey responses that identifies common challenges, recurring mistakes, exposure to different tools and frameworks, and takeaways for both students and educators.
\item Through course evaluation, we derived a set of recommendations for educators to inform the design and continuous improvement of similar courses in AI and software engineering education.

\end{itemize}
The rest of the paper is organized as follows:
Section~\ref{related_work} reviews the related work and situates this study within the existing literature. Section~\ref{Course_Design} describes the course design, learning outcomes, and assignments. Section~\ref{Methodology} outlines the research methodology and data analysis approach. Section~\ref{Result} presents the empirical results addressing the research questions. Section~\ref{Discussion} presents the discussion on meeting the course goals. Section~\ref{Threats} outlines threats to validity, and Section~\ref{Concluding} concludes the paper with final remarks and future directions.
\section{Related Work}
\label{related_work}
The growing use of AI in software-intensive systems has prompted researchers to highlight key differences between AI-enabled systems and traditional software. Early work by Sculley et al. \cite{sculley2015} introduced the concept of hidden technical debt in ML systems. They argued that many long-term problems do not stem mainly from inaccurate models, but from system-related issues such as complex data dependencies, difficult configurations, and tightly coupled pipelines. This view was later reinforced by Amershi et al. \cite{Amershi2019}, who showed that developing machine-learning-based systems requires careful attention to software engineering activities, including requirements, architecture, deployment, and maintenance throughout the system lifecycle. \\
The challenges of engineering AI-based systems have drawn increasing attention from software engineering educators. This has led a growing number of universities to start to integrate AI-based system engineering into their curricula. Kästner and Kang \cite{Kastner2020} present a master’s-level, project-based course that addresses the gap between model-centric AI education and the system-level challenges of building AI-enabled systems. The course follows an SE4AI perspective and focuses on requirements engineering, architectural decisions, and trade-off analysis. A simulated movie recommendation scenario is used to help students experience challenges such as data drift and feedback loops. Authors report on their reflections on course design instead of a systematic evaluation of student learning outcomes.

Xu et al. \cite{xu2025beyond} presented a course on self-adaptive systems that integrated hands-on learning with industry-relevant technologies to bridge the gap between academia and industry. To evaluate the learning outcomes of the course, they systematically analyzed student feedback. 
Lanubile et al. \cite{Lanubile2023} present a project-based university course that integrates machine learning and MLOps concepts to support end-to-end development workflows. Their results show that hands-on projects help students better understand the full development pipeline. However, the study also reports that model development remains the primary focus in many educational settings, while system-level aspects such as requirements evolution, architectural trade-offs, and system adaptability receive limited attention.

Mashkoor et al. \cite{Mashkoor2023} presented a graduate-level course titled ``Engineering of AI-Intensive Systems'' with project-based learning to bridge the gap between AI and software engineering students. The course was designed to bridge the gap between AI and SE students by working on collaborative project-based learning. Students from different disciplines teamed up to work on industry problems such as a gesture-control unit for medical assistance, an AI-based recipe finder, an image-generation app, and an AI-powered customer-support chatbot. The assessment consisted of a continuous evaluation of engineering documents (e.g., requirements and design specifications), final project demonstrations, and a written test. The findings showed that the interdisciplinary, experiential learning approach not only helped the students gain a more profound understanding of SE concepts in AI environments but also improved their hands-on problem-solving skills.

While existing studies offer important insights into course design and teaching practices for AI-enabled systems, student feedback is often limited or not systematically analyzed. Our work differs from prior studies \cite{Kastner2020} by combining the design of course content, assignments, and projects with a dedicated questionnaire to collect and analyze students’ feedback. This allows us to collect the students’ perspective on learning software engineering concepts in AI-intensive systems.
\section{Course Design}
\label{Course_Design} 
This section describes the design and delivery of the course \emph{AI Algorithms: Theory and Engineering}, focusing on its learning objectives, scaffolded assignments, semester-long project, and grading scheme.

\subsection{Course Description and Objectives} \label{sec:2.1-course_description}

The course is offered in the first semester of the international master's program in Artificial Intelligence and Intelligent Systems at the University of Bremen. As a mandatory element, the course adopts a project-based learning model that emphasizes practical and systemic challenges of engineering AI-enabled software systems within real-world constraints. In developing the course, we drew inspiration from Kästner's Engineering AI-Enabled Systems course\footnote{https://github.com/ckaestne/seai/}
at Carnegie Mellon University~\cite{kastner2025}, which demonstrates the value of confronting students with the gap between isolated ML models and full-scale AI-intensive systems. Our adaptation retains this focus but extends it by exposing students to contemporary technologies and architectural patterns (e.g., microservices, containerization, scalable deployment infrastructures), highlighting the interplay between AI components and their software ecosystems. 

The course targets students with prior training in Computer Science or related disciplines, such as Computer Engineering or Information Management. Admission requires demonstrated proficiency in programming, algorithms, and data structures, as well as prior experience with ML-based development, typically acquired during bachelor-level projects.
While most students enter with a solid understanding of ML methods and component-level software development, they often lack exposure to engineering constraints that arise when embedding learning-based functionality into large-scale systems. These constraints include ensuring robustness and interpretability of deployed components, managing operational costs, and maintaining data quality and runtime observability in dynamic environments. To address this gap, the course shifts focus from isolated ML development to system-level engineering of AI-enabled applications that must operate reliably under these demands.
The workload is defined as 56 hours of lecture time and 124 hours of self-guided study, totaling 180 hours and corresponding to six ECTS credits. The course has three principal learning outcomes.

\begin{tcolorbox}[
colback=blue!5,
  boxsep=0pt,
  left=1pt,right=1pt,
  top=3pt,bottom=3pt,
  before skip=2pt,
  after skip=2pt,  breakable
]
\textbf{Technical proficiency (L1):} \emph{Students gain the ability to interpret and address the requirements by applying advanced data structures, libraries, and frameworks to deploy AI components.} This competency is crucial because modern AI practice depends on algorithmic knowledge and effective integration of existing toolchains and frameworks.
\end{tcolorbox}

\begin{tcolorbox}[
colback=blue!5,
  boxsep=0pt,
  left=1pt,right=1pt,
  top=3pt,bottom=3pt,
  before skip=2pt,
  after skip=2pt, breakable
]
\textbf{Analytical competence (L2):} \emph{Students learn to evaluate the design and implementation trade-offs with respect to non-functional properties such as robustness, interpretability, and operational cost.} These considerations are essential for systems in which correctness cannot be reduced to accuracy metrics alone, but must account for operational constraints and societal expectations.
\end{tcolorbox}

\begin{tcolorbox}[
colback=blue!5,
  boxsep=0pt,
  left=1pt,right=1pt,
  top=3pt,bottom=3pt,
  before skip=2pt,
  after skip=2pt, breakable
]
\textbf{System-level understanding (L3):} \emph{Students acquire experience in designing fault-tolerant and scalable infrastructures for deploying AI-enabled systems.} They practice techniques for monitoring the runtime behavior, data quality, and provenance. This system-level competence reinforces the understanding that machine learning algorithms constitute integral components of broader system architectures rather than isolated technical artifacts.
\end{tcolorbox}

Recent empirical studies on ML-enabled software systems \cite{Sens2025}\cite{Serban2020}\cite{Tekinerdogan2025}\cite{Giray2021} underscore the \emph{relevance} of these learning outcomes, emphasizing that effective ML adoption requires a holistic, system-centric perspective beyond model development.

\subsection{Assignments and Project}
\label{assignments}

In this course, the registered 26 students are assigned to seven groups of three to six members to promote collaborative problem-solving. Five biweekly assignments are designed to progressively lead to a final project, all of which are submitted at team level. Assignments are report-based and code-based, where the former are considered for incremental grading and the latter are cumulatively graded at the end of the semester.

Weekly lab sessions are conducted to introduce assignments, present relevant tools, and address questions related to running assignments and the project. During these sessions, tutors also introduce the tools, technologies, and frameworks necessary for the effective implementation of the project. Following each assignment deadline, detailed feedback is provided through the submission repository.

The assignments and project are designed to jointly emphasize software engineering and machine learning aspects in the development of AI-enabled systems.

\textbf{The role of software engineering}: Students address software engineering challenges related to requirements engineering, architectural design decisions, trade-off analysis, telemetry integration, scalability, fault tolerance, and adaptation to evolving requirements.

\textbf{The role of machine learning}: Machine learning plays a central role in the development of a personalized movie recommendation system, requiring students to implement the complete ML workflow, including preprocessing, feature engineering, model training, evaluation, and model switching across recommendation approaches. Students experiment with algorithms such as Random Forest, Decision Trees, and K-Nearest Neighbors (KNN) integrated into collaborative and content-based filtering pipelines, while performing dataset cleaning, normalization, and train-test splitting to mitigate overfitting. These techniques are widely adopted in recommendation systems and are often complemented by context-aware recommendation strategies \cite{Habibi2022} to further improve personalization quality. Model selection and design choices are justified through trade-off analysis using metrics including accuracy, precision, recall, and F1-score.

The design and structure of each assignment, as well as the project, are described below.
\vspace{0.2cm}
\begin{sectionbox}
\textbf{\centering Assignment 1 -- Initial prototyping of AI-enabled recommendation system}

\end{sectionbox}
This assignment is a training exercise to understand the gap between the problem space and solution space. It is aligned with learning outcome \textbf{[L1]} and focuses on designing a modular pipeline for an ML based movie-recommendation system by defining system and feature goals, user goals, and model goals. Students integrate pre-trained \textit{ML algorithms} with publicly available \textit{datasets} with functionalities of data loading, pre-processing, user-profile generation, and \textit{feedback loop} integration to align with the defined goals. We emphasize \textit{system thinking} and the \textit{modularity} of the architecture in this assignment, with the goal of evaluating different models, analyzing their compatibility with datasets, and justifying the most suitable choices.
\vspace{0.2cm}
\begin{sectionbox}
\textbf{Assignment 2 -- Towards system designing via requirements engineering}

\end{sectionbox}
This assignment introduces the requirements engineering~\cite{kastner2025}, where students will identify \textit{goals} from the point of view of feature, user and model for developing scalable and secure recommendation systems. The world-machine-shared phenomena discussed in the lecture are used to identify the \textit{requirements}, \textit{specifications}, and \textit{assumptions} to realize these goals. This assignment aligns with the learning outcome \textbf{[L1]}, and it assists in the identification of system behavior leading to design decisions. 
\vspace{0.2cm}
\begin{sectionbox}
\textbf{Assignment 3 -- Architecture design decisions}
\end{sectionbox}
In this assignment, we give an initial idea of different aspects to consider in a software architecture design. Students will explore various solutions and justify which particular architectural design decisions are suitable for their project. This includes decisions regarding the selection of monolithic vs micro-service-based architecture, synchronous vs asynchronous communication, API-design strategies, client or server side inference location, telemetry data storage consideration, and centralized vs decentralized database design. Students will also design the control-flow and data-flow diagram in the deliverable. This allows students to explore different tools and methodologies that can assist in their projects. With this assignment, we can achieve both \textbf{[L2]} and \textbf{[L3]} learning outcomes. 
\vspace{0.2cm}
\begin{sectionbox}
\textbf{Assignment 4 -- Trade-off analysis}
\end{sectionbox}
To draw attention to the significance of systematic trade-off analysis, we introduce the \textit{Architecture Trade-off Analysis Method}~(ATAM)~\cite{kazman2000} through this assignment. The objective of ATAM is to foresee the consequences of design decisions based on the \textit{quality attributes} identified via requirements. This helps in deciding efficient allocation of resources and identifying decisions that require detailed analysis. 
This assignment involves identifying different quality attributes based on the \textit{scenarios}, which are elicited via requirements. These attributes represent the system's health and can include the performance, security, availability, and modifiability of the system. These attributes are further granulated as sub-factors into different levels and associated with the identified scenarios. Emphasis is placed on identifying \textit{measurable sub-factors}, which later help the trade-off analysis of design decisions. Each of these scenarios is associated either with qualitative or quantitative metrics in two dimensions: ease of implementation and significance to the overall success of the system. These scenarios are further used to identify design decisions required to realise them. Individual design decisions are associated with possible risks, sensitivity points, and trade-offs, which make use of the importance metrics assigned to the respective scenarios. This task is related to the learning outcome \textbf{[L2]}.
\vspace{0.2cm}
\begin{sectionbox}
\textbf{Assignment 5 -- Adapting to evolving requirements} ~\label{asst-5}

\label{assignment5} 

\end{sectionbox}

This assignment focuses on adapting the architecture to meet \textit{evolving requirements}. Students are provided with up to two team-specific and six common \textit{user stories} intended to modify their software architecture. Each team may select any four of the common user stories for implementation and explain the architectural changes along with their validation. These modifications reflect challenges inherent in developing AI-enabled systems, such as integrating heterogeneous datasets, switching or combining ML models, incorporating various learning strategies (e.g., incremental learning), real-time monitoring of telemetry data, and managing the versioning of learned models. The team-specific user stories introduce constraints on the common user story of integrating multiple ML models. The team-specific and common user stories together prompt the reconsideration of design decisions such as caching, telemetry data analysis, fault tolerance, and data distribution mechanisms. This assignment aligns well with the learning outcome \textbf{[L3]}.
\vspace{0.2cm}
\begin{sectionbox}
\textbf{Project}
\end{sectionbox}
In the project, students are tasked with developing an end-to-end movie recommendation system. The required deliverables include: (i) a concise report describing major system features, (ii) installation instructions accompanied by a video demonstration, and (iii) a version-controlled repository (Git/GitLab). Students are also encouraged to implement a basic user interface.
Project evaluation is based on predefined criteria covering both technical quality and system completeness. These include data integration and feature extraction, architectural design, advanced personalization, API and database design, telemetry monitoring, model retraining with feedback loops, overall pipeline integration, and code quality.
\subsection{Grading}

\textit{Individual grading}: Students are graded individually through an oral examination to balance teamwork and individual accountability. The oral exam assesses each student’s understanding of their project submission and its connection to lecture concepts. During the examination, students may be asked to explain specific parts of the codebase to evaluate both implementation details and overall project comprehension. \\
\textit{Team-wise assignment grading}: Assignments are graded on a team basis. Deliverables include pseudo-code, presentations, and documentation. Evaluation is based on evolving requirements, architectural design decisions, adaptation to changes, and justification of decisions, as described in the assignment description. Assessment considers completeness, alignment with specified deliverables, and adherence to fundamental concepts.
Project grading (Team-wise): The final project submission is graded using weighted evaluation criteria. These include submission quality aspects (code quality and documentation of instructions) and project-specific aspects (architectural design, documentation of design decisions, and the ML pipeline). Assignments collectively account for 50\% of the project grade, whereas the project account to the remaining 50\%.  \\
\textit{Bonus Grading}: In addition to formal assessments, active participation throughout the course is rewarded through bonus grades, according to the course assessment policy. \\
\textit{Grade Distribution}: The final course grade is determined by weighting the oral examination at 30\% and the project at 70\%.

\section{Research Methodology}
\label{Methodology}
\subsection{Data collection}
\subsubsection{Questionnaire}
To collect the necessary data required to answer our research questions, we designed a questionnaire which is divided into three main categories: students’ background in software engineering, their reflections on what they learned from the course, and their evaluations of the course from their own perspective. The estimated completion time is approximately 20 minutes. The questionnaire consists of 8 Likert-scale questions and 14 open-ended questions, which can be found along with student responses via Zenodo\footnote{https://zenodo.org/records/18417206}.

\textbf{Participants:}
The participants in the questionnaire were the students enrolled in the AI-Theory and Engineering master’s program offered during the winter semester 2024/25 at the University of Bremen. A total of 26 students registered for the course examination. {All 26 students were invited to participate in the anonymous survey on a voluntary basis with an option to skip any question they did not wish to answer. Of these, 14 students chose to participate.
\\
\subsubsection{Student assignment and project}
The artifacts consist of seven team-based submissions, but not individual submissions. It includes analytical reports, architectural designs, technical documentation, and implemented code produced across assignments and the project. The feedback provided to students on these artifacts are also taken into account.

\subsection{Data analysis}
\subsubsection{Qualitative analysis for RQ1 - RQ4}
In this study, we used the thematic analysis~\cite{Boyatzis1998} approach to analyze the data from the questionnaire by identifying and extracting common patterns and themes in participants’ responses. The data are first reviewed to understand their content. Open coding is then applied to label data with descriptive codes that capture meaningful units. These codes are grouped into broader themes, which are then reviewed and interpreted to obtain insight into the conceptual patterns.

\subsubsection{Qualitative analysis for assignments and project}

To answer RQ1.1, we conducted a qualitative analysis of the feedback provided on student-produced artifacts throughout the course, together with documented observations from teaching and tutoring activities.

\section{Result}
\label{Result}
\subsection{ Demographic Data of the Survey Participants }
A total of 14 students participated in the survey. The following section summarizes their demographic and background information based on six questionnaire items.\\
\textbf{Education:} All participants held a bachelor’s degree in computer science or computer engineering. Among them, two students specialized in artificial intelligence and data science, and one in robotics and intelligent systems.
\\
\textbf{Years of Experience as Software Developer before Taking the Course:} All 14 students responded to this question. The majority, six respondents (42.9\%), reported having no prior professional experience as software developers. The remaining respondents were evenly distributed, with two participants (14.3\%) each reporting 1, 2, 3, and 4 years of prior experience.
\\
\textbf{Software Architecture Design Experience:} Nine students provided examples of projects in which they had participated in software architecture design. The projects varied widely, including \textit{``Tourism Recommendation System''}, \textit{``Facial Recognition System''}, and many similar projects. We observed that a subset of the students demonstrated prior experience and familiarity with software architecture concepts. 
\\
\textbf{Software Project Development Experience:} Ten students reported having prior experience with learning-enabled projects, such as \textit{``Brain Tumor Segmentation...''}, \textit{``Stock Price Prediction...''}, and \textit{``Advanced Traffic monitoring system...''}. 
\\  
\textbf{Experience with Requirements Engineering:}
Thirteen students responded to this question. Of these, eight respondents (61.5\,\%) indicated having applied requirement engineering practices in one or more previous projects, whereas five (38.5\,\%) had not engaged in requirement engineering before this course.
\\  
\textbf{Tools and Frameworks Used Prior to the Course:} Thirteen responses were received for this item. The most frequently mentioned tools were Flask (7 times), Docker (5), and Python (4). Other commonly used technologies included PyTorch (3), AWS, PostgreSQL, Flutter, TensorFlow, Streamlit, MongoDB, and ReactJS (2 each). Several respondents also mentioned diverse frameworks and tools such as SQL, PowerBI, Node.js \& NPM, Django, Microsoft Visual C++, Kubernetes, Kafka, Heroku, FastAPI, and Vuforia Engine (1 each). The diversity of tools suggests that students had prior exposure to various software development environments and deployment platforms.

\subsection{Common Challenges from Students' Perspectives (\textbf{RQ1})}

To address this research question, we used qualitative data obtained from the questionnaire. By analyzing the responses to questions 8, 10, 12, and 14, we were able to answer this research question.

\begin{enumerate}
    \item \textbf{Team Collaboration and Knowledge Gaps:} A significant challenge cited across multiple phases was achieving effective team collaboration, which was often compounded by varying levels of technical expertise. Several respondents pointed to a lack of shared understanding and communication breakdowns as major hurdles. As one participant noted:
\begin{quote}
 \textit{``Most challenging aspect: Having a shared understanding with all of your teammates.''}\end{quote}
This was further exacerbated when team members made unilateral decisions, leading to friction and inefficiency. The knowledge gap was particularly acute in ML, where beginners struggled to keep pace, as expressed in this response: \begin{quote}
\textit{``Catching up in an environment where you have some colleagues already have knowledge in Machine Learning was a challenge.''} \end{quote}
This combination of communication issues and skill disparities made it difficult to establish a cohesive development process and make consistent architectural decisions.

    \item \textbf{Scalability and Adaptability of the System:}
A major challenge identified by participants was building a scalable and adaptable software system. Many teams struggled to make their solutions flexible enough to accommodate evolving requirements and future extensions. As one respondent noted:
\begin{quote}
\textit{``Making the software capable enough to accept and incorporate new changes easily.''}\end{quote}

    \item \textbf{Limited Prior Knowledge in ML:}
As highlighted by several respondents, limited prior experience with ML concepts and deployment technologies was one of the most persistent challenges. Many students noted that they had to quickly learn new tools and frameworks during the project, which affected their productivity and confidence. One student elaborated on:

\begin{quote}
 \textit{``The most challenging would be being a beginner in Machine Learning and the choice of models to choose for the project.''}
\end{quote}

Similarly, others emphasized the difficulty of adopting new technologies for integration and deployment. As one participant described:
\begin{quote}
 \textit{``It was a bit difficult to learn a new technology, for example Kafka, and then making sure it works fine.''}
\end{quote}

Such comments suggest that the steep learning curve associated with ML concepts and DevOps tools (e.g., Docker, Kafka, CI/CD pipelines) created additional overhead in both understanding and applying software engineering and ML principles. According to several respondents, this lack of familiarity also influenced their confidence in decision-making and slowed down experimentation and debugging processes.

    \item \textbf{Data Quality and Model Robustness:} From a pure ML perspective, the primary challenge shifted from theoretical model selection to the practical difficulties of data preparation and achieving robust performance. Participants consistently reported that real-world data was a major obstacle. Issues such as missing values, noisy inputs, and imbalanced datasets made it difficult to train reliable models. One response clearly articulated this:
\begin{quote}
 \textit{``The most challenging part of the project was handling real-world data. Dealing with missing values, imbalanced datasets, and noisy inputs made it difficult to achieve consistent model performance.''}\end{quote}

    \item \textbf{Complexity of Hybrid Architectures:} Some projects involved implementing hybrid recommender systems or combining multiple models. One participant described: \begin{quote}
 \textit{``Implementing a hybrid recommender system was more challenging than implementing a pipeline with just one model.''}\end{quote}

    \item \textbf{Technical Integration:}
Connecting multiple modules and ensuring seamless communication among them was a recurring difficulty. Respondents mentioned that the integration phase often exposed unforeseen issues in scalability and system performance:
\begin{quote}
 \textit{``Connecting different modules to each other.''}\end{quote}
\begin{quote}
 \textit{``Ensuring seamless model deployment, maintaining real-time performance and scalability.''}\end{quote}

    \item \textbf{Design Overhaul and Data Re-engineering:} Many participants shared that changes in project requirements forced them to make major updates to their systems. These included adding new models, updating databases, retraining existing models, and even redesigning parts of the user interface. For example, one participant said,

\begin{quote}
 \textit{``We needed to add another model and it changed our DBs, inputs, model architecture, UIs, etc.''}\end{quote}
 Others mentioned the effect of integrating new dataset as,
\begin{quote}
 \textit{``Updating the database with item meta (directors, crew, and cast for movies) and retraining the model.''}\end{quote}
In some cases, effects of changes in performance metrics were noted as,
\begin{quote}
 \textit{``We set a different requirement for model accuracy which required a change of the whole ML component design.''}\end{quote}
 Overall, these experiences show that minor changes in requirements can have a big impact across multiple parts of a system.

    \item \textbf{Team Dynamics and Project Management:} Beyond technical challenges, many participants found that team and organizational issues had a big impact on the project and team morale. Managing disengaged or unsupportive team members often put extra pressure on motivated individuals. One participant explained, \begin{quote}
\textit{``My team was not supportive (They did not care much about the project), so I had to handle it all by myself.''}\end{quote}
Communication and decision-making were also difficult in some teams. Reaching consensus was sometimes challenging, and in a few cases, biases affected how ideas were accepted: \begin{quote}
\textit{``The project was submitted successfully, but biased acceptance of ideas based on ethnicity hindered creativity, and excluding valuable input as well as understanding of project.''}\end{quote}
These experiences show that strong team management and an inclusive environment are essential for success in complex, collaborative projects, alongside technical skills.

\end{enumerate}

\subsection{Common Challenges from Educators' Perspectives (\textbf{RQ1.1})}

By reviewing the students’ projects and their assignments, we identified the following challenges from our perspective.

\subsubsection{Educational Challenges}
\begin{itemize}[left=0pt]
\item \textit{Teaching:} While it is relatively straightforward to differentiate between requirements, specifications, and assumptions, effectively conveying appropriate abstraction levels across them remains difficult, as interpretations vary by domain, prior knowledge, and project scope. Additionally, practical engagement with the ATAM method necessitates numerous revisions and their subsequent tracking, particularly when adapting to AI-enabled systems.
\item \textit{Knowledge gap:} Disparities in exposure to software engineering principles and ML tools result in inconsistent engagement in the course, which can be addressed by adaptable tutorial support.
\item \textit{Organizational and infrastructure:} Students switching teams or withdrawing mid-course, limited access to computing resources, and inconsistent attendance should be taken into account before course design.
\end{itemize}

\subsubsection{Technical and Architectural Challenges}

\begin{itemize}[left=0pt]

\item \textit{Model switching mechanisms:} Students frequently struggled to design dynamic model selection strategies, largely due to limited architectural modularity, inadequate anticipation of performance variability, and insufficient planning for monitoring and observability.
\item \textit{Fallback strategies:} Difficulties in implementing robust fallback models were common, primarily due to weak modular design practices and limited consideration of model versioning and degradation scenarios.
\item \textit{User profiling design:} Difficulties with long-term and short-term user profiles were primarily caused by the inherent complexity of simulating evolving user behavior and insufficient preparation of datasets to capture these dynamics.
\item \textit{Dependency management:} Challenges in managing software dependencies were primarily rooted in limited prior exposure to professional development workflows and integration practices, which is common among early-stage graduate students.

\end{itemize}

\subsection{Effectiveness of Bridging the Gap Between Theory and Practice (\textbf{RQ2})}

There exist well-recognized challenges in integrating ML components into software systems~\cite{Sens2025}~\cite{Paleyes2022}~\cite{Serban2022}, best understood through project-based approaches complementing theoretical teaching.  While theory builds foundational understanding, projects enable work with real-world data and system integration challenges. These challenges are difficult to anticipate due to problem nature, datasets, technology, and students' experience levels.

Student responses reflected these observations. Participant noted that earlier ML courses focused on models rather than complete systems, with one student remarking,\textit{``Machine learning is not about constructing models; it’s about data and preprocessing''}. Others described difficulties in learning and integrating several frameworks and tools.

The course applied theoretical concepts taught in the lecture through a semester-long project. Tutorials provided hands-on exposure to tools including including Kafka for data streaming, Flask for microservice APIs, Prometheus and Grafana for monitoring, Locust for load testing, Gunicorn as the microservice server, and Nginx for load balancing. Assignments reinforced incremental development along with team-specific evolving requirements that intentionally challenged students' design decisions, allowing practical application and exploration of additional tools.

The questionnaire responses demonstrate how students navigated this transition from theory to practical implementation. Their reflections on the course structure, the challenges, and the strategies they employed to handle evolving requirements provide direct evidence relevant to this research question. Both qualitative and quantitative analyses were used to evaluate the course's effectiveness in bridging theory and practice via practical tools, experience with adapting to evolving requirements, and engagement with system-level integration. These insights are summarized in the themes below.
\begin{enumerate}[left=0pt]
    \item \textbf{Tool-driven transition from theory to practice:} The course structure provided sustained exposure to practical technologies, enabling students to bridge theoretical understanding with real-world implementation skills. Questionnaire responses show that students worked with a diverse set of tools throughout the semester, as illustrated in Figure~\ref{fig:tool_usage}. Frequently mentioned tools included Docker (7 participants), Kafka (6 participants), Flask (5 participants), and MongoDB (5 participants). Others reported using Kubernetes (3 participants), PyTorch (3 participants), Prometheus (2 participants), Grafana (2 participants), and React (2 participants). A subset of students explored more advanced frameworks such as FAISS, ZMQ, PyQT, Locust, Nginx, Redis, and even tools such as MLFlow. Overall, the course supported both the deepening of pre-existing tool knowledge and the exploring technologies such as Waitress, Maven, and the OTL–LGTM stack, thereby facilitating a hands-on transition from conceptual understanding to implementation in a realistic engineering environment.

    \item \textbf{Learning through evolving requirements:} As a result of adapting requirements from Assignment 5, students were asked to modify their software via user stories. These changes were deliberately designed to simulate realistic system-driven requirement evolution as discussed under Assignment~5, which was also influenced by architectural weaknesses identified in earlier submissions. Students were informed during Assignment 4 that requirement changes would occur but were not told their specific nature, scope, or timing, creating bounded uncertainty. This allowed preparation at an architectural level without enabling tailoring designs to exact modifications. Students consistently reported that evolving requirements prompted concrete changes at multiple levels of their systems, including architecture, ML model design, database structures, and the integration of additional security or monitoring components. One team described how a single requirement change influenced decisions across ML models, API design, and data-flow pipelines. Others highlighted the difficulties of merging newly introduced datasets with their existing data structures, leading to substantial downstream modifications in their codebase. One student summarized this as, \textit{``making software capable of accepting new changes easily was the biggest lesson''}. However, some teams had minimal changes as a result of initial microservice-based architecture and modularity in the codebase. 
    
    These experiences demonstrate that the course successfully simulated real-world uncertainty in AI-enabled system development. By encountering requirement shifts, sometimes late in the development process, students were required to engage in architectural revision, trade-off evaluation, and iterative refinement. Such forms of experiential learning represent an essential aspect of practical implementation that cannot be gained through theoretical instruction alone.

    Although no explicit reflection across teams was organized, the final presentations, held in a shared session, enabled implicit exchange on how requirement changes were addressed.

    \item \textbf{Integration complexity as practical learning:} A recurring theme across the responses was that students came to understand that integrating ML algorithms into a functioning software system is a complex, hands-on process that only becomes fully visible during implementation. Several participants noted that their earlier coursework focused narrowly on model training, whereas this project exposed them to the broader system-level challenges of real AI application development. As one student expressed, \textit{``Software systems with an ML core are more than just training a model''}. 

    Students detailed a wide range of integration-related challenges, including combining ML components with databases, APIs, microservices, and external services; managing deployment pipelines; instrumenting monitoring tools; and ensuring reliable performance under realistic workloads. Many also emphasized interpersonal and organizational challenges such as achieving shared understanding within the team or dealing with inconsistencies between modules developed by different group members, which further complicated integration work.

    Taken together, these responses indicate that integration complexity functioned not as a barrier but as a meaningful learning opportunity. The course enabled students to directly experience the practical realities of end-to-end AI system development, moving them beyond theoretical knowledge toward the applied competencies required in professional practice.
\end{enumerate}

\begin{figure}[tbp]
   \centering
   \includegraphics[
    width=\columnwidth,
    height=0.93\textheight,
    keepaspectratio
]{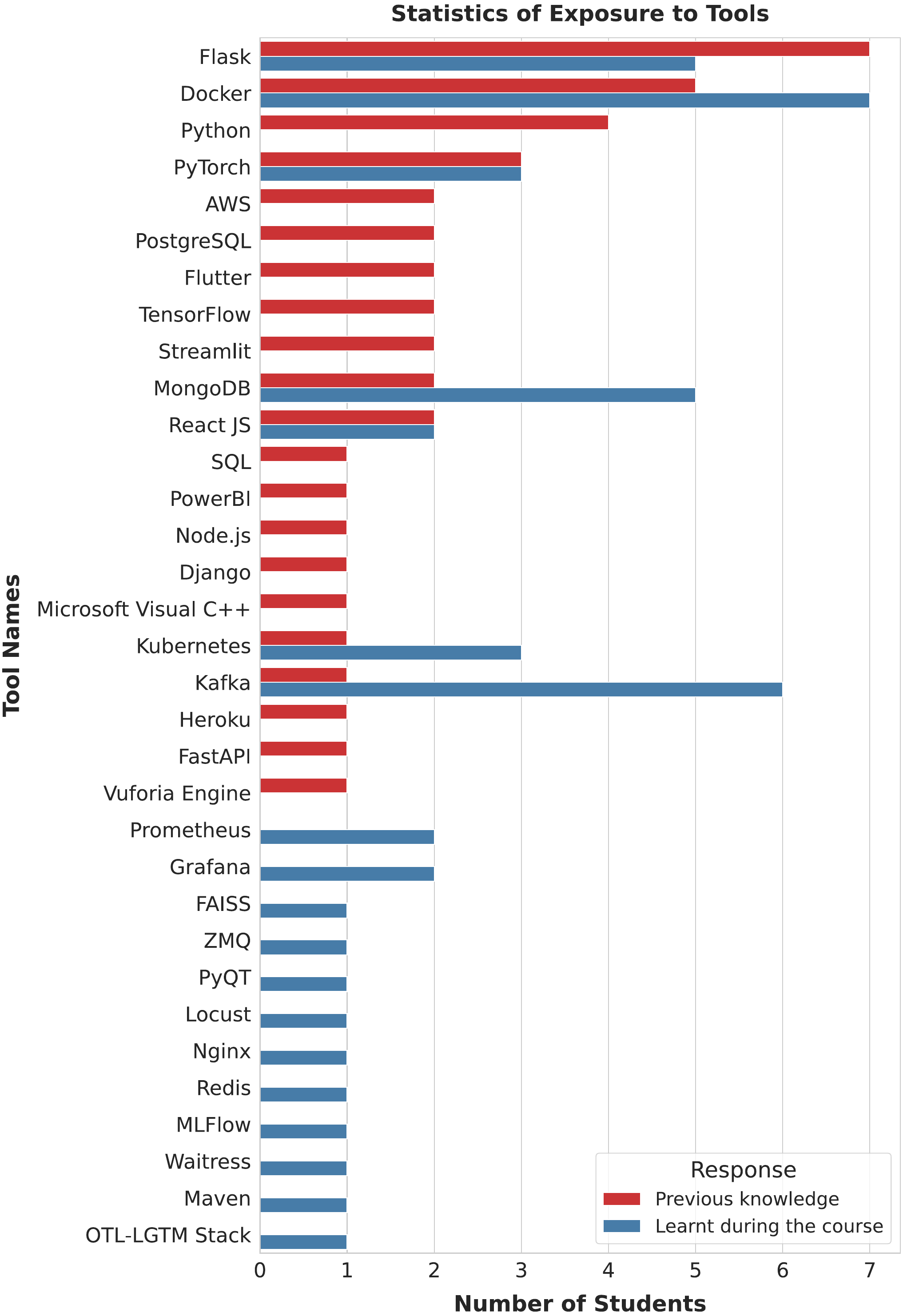}
   \caption{Distribution of student-reported tools and technologies used in the project.}
   \label{fig:tool_usage}
\end{figure}

\begin{figure*}[t]
    \centering
    \includegraphics[
        width=0.9\textwidth
    ]{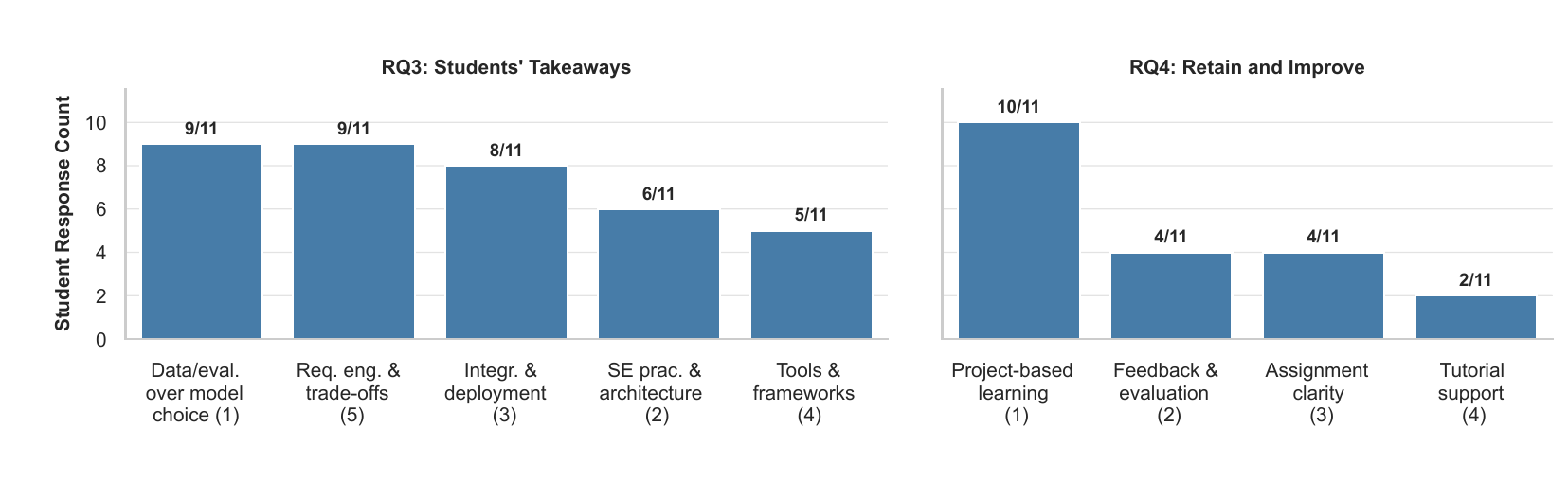}
    \caption{Theme frequencies for RQ3 and RQ4: Bars show the number of student responses mentioning each theme. Numbers in parentheses correspond to the numbered themes in Sections 5.5 and 5.6. Multiple themes could be coded in a single student response; counts are therefore not mutually exclusive.}
    \label{fig:rq_themes}
\end{figure*}

\subsection{Students’ Takeaways from the Course (\textbf{RQ3})}

The primary takeaway reported by students reflects a shift from viewing ML algorithms as an isolated component toward understanding it as an integral part of a larger AI-enabled systems. Other key insights include recognizing the critical role of data quality, preprocessing, and evaluation in determining ML performance. Equally significant was the appreciation of scalable, modular, and maintainable software architectures, supported by structured project management and software engineering practices such as requirements engineering, ATAM, and systematic testing.

The key themes identified from students’ responses to the questionnaire regarding takeaways in the perspective of software engineering, ML, and their integration are visualized in Figure~\ref{fig:rq_themes} and summarized below:

\begin{enumerate}[left=0pt] 
    \item \textbf{Data quality and evaluation outweigh model choice}: Nine student responses highlighted a recurring theme that data quality and evaluation practices had a greater impact on ML success than the choice of algorithms. Analysis of the open-coding process revealed that students consistently emphasized data preprocessing, dataset suitability, and evaluation strategies as central challenges. One participant noted:

    \begin{quote}
        ``\textit{Getting to train the models..., and then get proper datasets, and also have proper evaluation metrics to evaluate them. Making sure the models can work correctly in the production setting rather than only on the test set (data drifts, concept drifts)}''
    \end{quote}

    Several students also reported a shift in perspective, recognizing that effective ML development is fundamentally data-driven rather than model-centric. As one respondent summarized:

    \begin{quote}
        ``\textit{The greatest thing to take away from this course is realizing that machine learning is not about constructing models; it’s about data}''
    \end{quote}

    These responses indicate that students developed an understanding of ML systems in which data preparation and evaluation play a dominant role relative to algorithmic complexity.
    
    \item \textbf{Emphasis on best software engineering practices and designing scalable architectures}: Six student responses reflected a shared emphasis on the importance of applying established software engineering practices when developing AI-based systems. Students highlighted modular design and structured development processes as essential for achieving scalability, maintainability, and long-term adaptability. In particular, respondents recognized that disciplined engineering approaches were necessary to manage system evolution and operational complexity. As one participant stated:
    
    \begin{quote}
        ``\textit{Structured development processes like agile methodologies and testing frameworks help build scalable, maintainable software}''
    \end{quote}

    These responses suggest that students viewed AI development not merely as model construction, but as a software engineering activity requiring architectural foresight and systematic development practices.
    
    \item \textbf{Readiness for integration and deployment of ML components}: Eight student responses emphasized that ML models constitute only one part of a complete software system and must be supported by appropriate integration and deployment practices. Students highlighted the need for system-level design, deployment pipelines, and continuous integration mechanisms to ensure reliable operation beyond experimental settings. As one participant observed:

    \begin{quote}
        ``\textit{Deploying models into an efficiently laid out software framework is equally significant as accuracy}''
    \end{quote}

    Overall, these responses indicate that students developed an awareness of ML as an integrated software engineering activity, where deployment readiness and architectural alignment are critical for sustainable system performance.
    
    \item \textbf{Growth in technical tool proficiency and framework literacy}: Five student responses highlighted increased proficiency with development tools and ML frameworks as a key learning outcome. Students emphasized that exposure to a diverse toolchain enhanced their technical competence and preparedness for real-world system development, particularly in integrating multiple technologies into a coherent solution. One participant described this challenge as follows:

    \begin{quote}
        ``\textit{Learning new tools and then making all of them work together. Making sure that we are choosing the right technologies and that they can work in a coherent manner}''
    \end{quote}

    These responses suggest that students developed practical literacy in contemporary development ecosystems rather than isolated familiarity with individual tools.

    \item \textbf{Increased awareness of planning, requirement engineering, and design trade-offs}: Nine student responses reflected a growing appreciation for early planning, requirements engineering, and architectural trade-off analysis in AI system development. Students emphasized the importance of considering requirements and design decisions upfront rather than treating them as secondary to model development. This qualitative insight is reinforced by questionnaire data (Q13), where 75\% of students (9 out of 11) acknowledged the significance of ATAM in guiding architectural decisions. One participant noted:

    \begin{quote}
        ``\textit{Plan the requirements of a model that you are working with ahead and be well prepared for it}
    \end{quote}

    More broadly, students recognized AI system development as a multi-dimensional design activity involving requirements, architecture, model selection, and trade-offs across competing quality attributes. These responses indicate an increased awareness of structured decision-making processes in the design of complex, evolving systems.
\end{enumerate}
The aforementioned themes effectively support the realization of the learning outcomes outlined in course description. The first learning outcome, \textit{technical proficiency}, is reinforced by themes 2 and 4. The second learning outcome, \textit{analytical competence}, is supported by themes 2 and 5. The third learning outcome, \textit{system-level understanding} is supported by themes 1, 3, and 5. Together, these themes effectively realize the course's key learning outcomes.

\subsection{Areas for Course Improvement (\textbf{RQ4})}

The students appreciated the project-based, practical structure of the course, viewing it as the key driver of learning. However, they identified several improvement areas, including the need for more timely and constructive feedback with clearer assignment objectives. Suggestions also included adding technical workshops to strengthen implementation skills. The themes that addresses this research question are identified from survey question 21, and they are visualized in Figure~\ref{fig:rq_themes} and listed below along with takeaways from the educator's perspective:

\begin{enumerate}[left=0pt]
    \item \textbf{Keep practical, project-based learning}: Ten student responses consistently emphasized the effectiveness of the course’s practical, project-based structure. Students reported that hands-on assignments, iterative project development, and course alignment around a central project significantly enhanced their learning and understanding of real-world system design. One participant summarized this sentiment succinctly:

    \begin{quote}
        ``\textit{I think having a project that is the focus of the course is really interesting and made me learn a lot}''
    \end{quote}

    Overall, these responses indicate strong student support for retaining a project-oriented pedagogy that emphasizes practical engagement and real-world application as the primary learning mechanism.

    \item \textbf{Improve feedback and evaluation mechanisms}: Four student responses pointed to limitations in the timing and quality of feedback throughout the course. Students reported that feedback was often delayed or insufficiently detailed, which reduced its usefulness for guiding design decisions and correcting issues early in the development process. One participant noted:

    \begin{quote}
        ``\textit{We did not get feedback in a timely manner}''
    \end{quote}

    Several responses also emphasized the value of earlier and more challenging feedback, particularly during system design phases, to prevent the propagation of suboptimal architectural decisions. As one student reflected, early critique could have enabled course correction before significant effort was invested. Collectively, these responses indicate a need for more timely and formative evaluation mechanisms to better support iterative learning.

    \item \textbf{Clarity in scope of assignments}: Four student responses highlighted challenges related to the clarity and specificity of assignment instructions. Students reported that some tasks were perceived as vague, leading to uncertainty about expected outcomes, assessment criteria, and appropriate levels of effort. One participant noted:
    \begin{quote}
    ``\textit{especially assignment 2 and 3, I was confused and things were vague. you may invest more on teaching deeper concepts related to assignments 2 and 3}''
    \end{quote}

    Several responses emphasized the need for clearer objectives and more explicit requirements to better align student work with course expectations and evaluation metrics. Collectively, these responses suggest that while the assignments were grounded in realistic problem settings, clearer scoping and articulation of requirements would better support student understanding and execution.
    
    \item \textbf{Tutorial sessions directed towards assignment}: Two student responses highlighted the need for tutorial sessions that are more closely aligned with upcoming assignments. Students expressed a desire for targeted technical support, particularly in the form of demonstrations and hands-on guidance for tools and technologies used in the projects. One participant noted:

    \begin{quote}
        ``\textit{More workshops on the usage of technologies such as docker, flask etc. How they should be integrated together, how to use them with the current scenario}''
    \end{quote}

    Overall, these responses suggest that more assignment-focused tutorials could better support students in applying tools effectively within the project context.
    
    \item \textbf{Project and documentation should go hand-in-hand}: Although this theme is supported by a single student response, it raises a relevant concern regarding the alignment between documentation-oriented assignments and incremental system development. The respondent observed that several intermediate assignments focused primarily on theoretical aspects, which deferred implementation work and concentrated development effort toward the end of the course. As the student noted:
    \begin{quote}
        ``\textit{The 2nd, 3rd and 4th assignment was mostly theoretical… It didn’t force us to build/code anything, which meant that students had a lot of load at the very end of the project}''
    \end{quote}

    This response suggests that closer coupling of documentation activities with incremental implementation could help distribute workload more evenly and reinforce the connection between design artifacts and executable systems.

\end{enumerate}

\section{Discussion}
\label{Discussion}

Based on our experience of designing and teaching this project-based course and the findings from the Result section, this section discusses key takeaways which might be helpful for the readers incorporating similar courses focused on AI-enabled systems.

Throughout the course, students developed skills in setting goals and deriving requirements through Assignments 1 and 2, although some teams confused the concepts of goals, requirements, specifications, and assumptions. In Assignments 3 and 4, most groups demonstrated a good understanding of system qualities such as performance, scalability, and modifiability, but often overlooked aspects like maintainability, privacy, and user experience, and design was less connected to the actual building process. In Assignment 5, students had to revise their architectural decisions, highlighting the evolving nature of software design. 

Regarding the final project, many teams delivered usable and visually appealing UIs, and some implemented advanced technologies like microservices, security tokens, Grafana monitoring, Kafka, and Docker, showing a system-level mindset. Students, however, faced challenges with module integration, real-world deployment, and learning new tools.

In terms of the course goal, students gained a clear understanding that developing intelligent systems involves more than selecting models; it requires robust, scalable, and maintainable architectures. Many students emphasized that the most valuable aspect of the course was understanding the link between software architecture and ML and familiarizing themselves with engineering tools such as Docker, Kafka, CI/CD, and ATAM.

\section{Threats to Validity} 
\label{Threats}
\textbf{Internal validity:} 
    This study relies on qualitative analysis of questionnaire responses and student artifacts, which may be subject to researcher interpretation bias. To mitigate this threat, multiple data sources, including anonymous surveys and actual student submissions, were used.\\
   \textbf{External validity:} The conclusions drawn are from only one course at a single university with a small number of students, resulting in limitations of the generalizability of the findings. Nevertheless, the study yielded a number of meaningful insights, challenges, and recurring patterns. Accordingly, the results may inform the design and improvement of similar courses in the context of engineering AI-enabled systems, although direct generalization should be made with caution.\\
  \textbf{Construct Validity:} Threats to construct validity refer chiefly to how well the concepts being studied are reflected by the instruments used for their measurement. For instance, in this particular study, concepts such as learning difficulties, typical mistakes, and educational achievements were measured by combining students’ self-reported questionnaire responses with the analysis of project deliverables. However, the students answering the questionnaire might have been guided by their individual perceptions.
To mitigate this threat, a multi-source strategy was implemented, and the questionnaire data were verified by the objective evidence taken from the assignments and final project.
Learning difficulties, typical mistakes, and educational achievements were measured using students’ self-reported questionnaires and project deliverables. Since questionnaire responses could be influenced by personal perceptions, a multi-source strategy was applied by verifying them with objective evidence from their artifacts.

\section{Conclusions} 
\label{Concluding}

This paper examines how the project-based software engineering for AI-enabled system (SE4AI) course facilitated students to develop a real-world AI-enabled system by addressing four research questions on challenges, theory-practice integration, learning outcomes, and course improvement. Structured around clear learning outcomes, the course involved five scaffolded assignments and a semester-long project to build a machine-learning-based movie recommendation system covering requirement engineering, architecture design, ATAM, deployment, monitoring, and adaptation to evolving requirements.

A mixed-methods analysis of assignment artifacts, project submission, and surveys from 16 students reveals that they developed technical proficiency through the exploration and integration of frameworks and tools, analytical skills via architectural decisions and trade-off analysis, and system-level understanding through hands-on experience with distributed architectures and adapting architecture to evolving requirements.

Overall, the course effectively met learning goals and bridged gap between theory to practical system engineering, while identifying concrete directions for future improvement.

\section*{Acknowledgment}
The authors would like to express their sincere gratitude to all the students who generously contributed to this research by completing the questionnaire.

\clearpage
\end{document}